\documentclass[twocolumn]
{aastex631}
\usepackage{amsmath}
\usepackage{amssymb}
\usepackage{aas_macros}
\usepackage{appendix}
\usepackage{gensymb}

\begin{document}

\title{Geminga: A Window into the Role Played by the Local Halo in the Cosmic Ray Propagation Process}

\correspondingauthor{Yu-Hai Ge}
\email{geyh@ihep.ac.cn}
\correspondingauthor{Yi-Qing Guo}
\email{guoyq@ihep.ac.cn}
\author{Lin Nie}
\affiliation{School of Mechanics and Aerospace Engineering, Southwest Jiaotong University, Chengdu, 610031, China}
\affiliation{School of Physical Science and Technology, Southwest Jiaotong University, Chengdu, 610031, China}

\affiliation{Key Laboratory of Particle Astrophysics, Institute of High Energy Physics, Chinese Academy of Sciences, Beijing, 100049, China
}

\author{Yu-Hai Ge}
\affiliation{Key Laboratory of Particle Astrophysics, Institute of High Energy Physics, Chinese Academy of Sciences, Beijing, 100049, China
}

\author{Yi-Qing Guo}
\affiliation{Key Laboratory of Particle Astrophysics, Institute of High Energy Physics, Chinese Academy of Sciences, Beijing, 100049, China
}
\affiliation{University of Chinese Academy of Sciences, Beijing 100049, China}
\affiliation{TIANFU Cosmic Ray Research Center, Chengdu 610000, China}
\author{Si-Ming Liu}
\affiliation{School of Physical Science and Technology, Southwest Jiaotong University, Chengdu, 610031, China}

\begin{abstract} 
A novel phenomenon among the recently observed Geminga pulsar halo is the presence of distinct radiation morphology at high energies, while no extended radiation is detected in the 10-500 GeV energy band within a $40\degree\times40\degree$ region. This phenomenon suggests that pulsar halos play a crucial role in the local propagation of cosmic rays, making it necessary to investigate the underlying mechanisms of this phenomenon. This work focuses on the 3D propagation study of cosmic rays, incorporating the Geminga pulsar into our propagation framework to investigate its contribution to different observational spectra. We consider Geminga a dominant local source of positrons, partially reproducing the observed positron spectrum and multi-wavelength radiative spectra of the Geminga halo. Through calculations of signal and background at different angles, we find that: (1) The slow-diffusion properties near the Geminga pulsar and its proper motion may cause the radiation from electrons originating from Geminga to be distributed across a more extended region. (2) The incomplete subtraction of radiation from the local halo may contribute, to some extent, to the diffuse gamma ray fluctuations detected by LHAASO. We hope that LHAASO will detect more sources of cosmic ray halo to further validate our model.

\end{abstract}


\section{Introduction} \label{sec:intro}
As cosmic ray observation experiments enter an era of precision, traditional cosmic ray propagation models cannot predict some finer spectral structures. For instance, primary CR spectra exhibit a spectral hardening at high energies. Data from AMS-02 reveals that the CR spectra above several hundred GeV deviate from a simple power law, displaying a spectral hardening, a phenomenon now confirmed across various experimental platforms \citep{2021PhR...894....1A,2019SciA....5.3793A,2018JETPL.108....5A}. Secondary CR spectra also harden in this energy range, and their spectra are steeper than those of primary CRs beyond this energy. The latest observations by DAMPE further confirm an excess in the secondary-to-primary flux ratio of CRs at high energies \citep{2022SciBu..67.2162D}.

To explain these observations, a spatially dependent CR propagation model has been developed. In contrast to conventional models that assume uniform propagation of CRs throughout the Galaxy, we propose the following: (1) CR propagation depends on the local distribution of matter \citep{2018PhRvD..97f3008G,2016ApJ...819...54G,2012PhRvL.108u1102E}; (2) local CR sources act as both a potential well, confining local CRs, and a barrier, impeding external CRs from penetrating these regions \citep{2024PhRvD.109f3001Y,2024ApJ...974..276N,2012ApJ...752L..13T}; and (3) local CRs primarily consist of two components—component A, comprising CRs accelerated from distant sources and diffused into the region, which dominates at low energies, and component B, consisting of CRs from local sources, which dominate at high energies \citep{2024PhRvD.109f3001Y,2024ApJ...974..276N}. High-energy phenomena in the Galactic vicinity are thus predominantly driven by local sources. This propagation model effectively accounts for the observed CR spectral features. Besides, observations of diffuse gamma-ray emissions in the Galaxy and extended emissions from local CR sources may provide further validation of this propagation framework.

Gamma-ray extended emissions observed near pulsars confirm the existence of nearby regions with slower diffusion relative to the interstellar medium (ISM). Since the morphology of pulsar halos follows the spatial distribution of their parent electrons/positrons, they serve as ideal probes for studying local CR propagation in the Galaxy. Observations of extended radiation from pulsars at different energy bands provide an opportunity to test our model. Recently, the Milagro Collaboration \citep{2009ApJ...700L.127A}, High-Altitude Water Cherenkov Observatory (HAWC) \citep{2017Sci...358..911A}, HESS \citep{2023A&A...673A.148H}, and Fermi-LAT \citep{2019PhRvD.100l3015D} have all revealed radiation from the Geminga halo. Notably, the Milagro Collaboration reported gamma-ray radiation from the Geminga direction extending over 2 degrees and spanning the energy range of 1-100 TeV. This observation was confirmed by HAWC, which detected extended gamma-ray emissions coincident with the position of Geminga. However, a decade-long analysis of Fermi-LAT data indicates no significant extended emission around the Geminga region in the 10-500 GeV energy range within a $40\degree\times40\degree$ region \citep{2019ApJ...878..124Q,2019ApJ...878..104X}. Therefore, we believe that local CR halos play a crucial role in the propagation of CRs in the Galaxy, and investigating the mechanisms that generate the observability of pulsar extended emissions at different energy bands is essential.

This work focuses on the observations of radiation spectra in different energy bands from the Geminga halo, detected by Fermi-LAT and HAWC, as well as the distribution of diffuse gamma rays near Geminga observed by LHAASO, and the measurement of cosmic-ray positrons by AMS-02. These observations constrain the injection and diffusion of cosmic-ray particles in the Geminga halo and allow us to investigate the properties of extended radiation from the Geminga halo in different energy ranges. We aim to test the proposed spatially dependent cosmic-ray propagation model and physical framework. The structure of this paper is as follows: Section \ref{sec:method} introduces the diffusion of electrons/positrons in the pulsar halo; Section \ref{sec:result} discusses our results, and Section \ref{sec:conclusion} provides a summary and outlook for future work.

\section{Diffusion in Halo} \label{sec:method}
The local halo plays an essential role in the cosmic ray propagation process. The fluctuating structure observed in the radial distribution of diffuse gamma rays by LHAASO may be due to the contribution from local sources that cannot be fully subtracted, particularly in the regions near the Geminga and Cygnus bubbles \citep{2024ApJ...974..276N}. Therefore, we first use the spatially dependent propagation model proposed in previous work (see \ref{appendix}) to simulate the distribution of Galactic background cosmic rays and the corresponding background diffuse gamma-ray radiation that they produce. Then, we consider the halo as a slow diffusion region and, 
calculate its radiation spectrum to explain the latest observational data from LHAASO, Fermi-LAT, and HAWC. Finally, we discuss the properties of the Geminga halo signal under the noise of the background radiation and the proper motion of Geminga pulsar.

The propagation equation for electrons in the Geminga halo is the same as the propagation equation \ref{eq3} for background cosmic rays. The Geminga halo is treated as a individual cosmic ray source within the Galprop framework. Since diffusion in the halo differs from that in the general ISM environment, diffusion within the Geminga halo is considered to occur in a slow diffusion region. Once the electrons escape from the Geminga halo, their diffusion follows the same scenario as the background cosmic rays.
the electrons from Geminga are continuously injected halo around it, and the injection spectrum is normalized so that the total power injected is given by the expression\citep{2019ApJ...879...91J} $\rm L(t)=\eta \dot{E}_0\left(1+\frac{t}{\tau_0}\right)^{-2}$.  Here, $\rm E_0$ is the initial spin-down power of the pulsar and $\rm \tau_0=13~kyr$, the initial spin-down power is calculated using the current spin-down power of $\rm E=3.26\times 10^{34}~erg s^{-1}$ assuming that the pulsar age is $\rm T_{age}=340 ~kyr$.

The diffusion coefficient around the pulsar, estimated from gamma-ray observations, suggests that it may be several hundred times smaller than the average value across the entire Milky Way. This slow diffusion could be caused by electrons/positrons escaping from the pulsar itself or by turbulent regions generated by cosmic ray particles escaping from the pulsar’s progenitor supernova remnant. Therefore, in our current work, we treat the diffusion of cosmic ray particles from the pulsar halo as a spherically symmetric ``two-zone'' diffusion model, similar to the scenario provided in \cite{2018ApJ...863...30F}, which consists of slow diffusion within the halo and faster diffusion in the interstellar medium. The diffusion coefficient is mathematically described as \citep{2019ApJ...879...91J}:
\begin{equation}
D=\beta\left(\frac{\mathcal{R}}{\mathcal{R}_0}\right)^{\delta_h} \begin{cases}D_h, & r<r_i \\ D_h\left[\frac{D_0}{D_h}\right]^{\frac{r-r_i}{r_o-r_i}}, & r_i \leqslant r \leqslant r_o,\end{cases}
\label{eq1}
\end{equation}
where $\rm \beta$ represents the particle velocity in units of the speed of light, $\rm D_0$ is the normalization of the diffusion coefficient in the general ISM, $\rm D_h$ is the normalization for the diffusion coefficient within the SDZ with radius $\rm r_i$, $\rm R$ is the particle rigidity, and $\rm R_0=4~GV$ is the normalization (reference) rigidity. The zone between $\rm r_i$ and $\rm r_o$ is a transition layer where the normalization of the diffusion coefficient increases exponentially from $\rm D_h$ to the ISM value $\rm D_0$. For the slow-diffusion region of the Geminga halo, we fix the diffusion coefficient index $\rm \delta_h$ to 1/3. The diffusion coefficient of electrons in the halo is described by Equation \ref{eq1}, while after they enter the ISM, their diffusion coefficient is described by Equation \ref{eq4} in the \ref{appendix}.

The electrons/positrons injected into the halo by the pulsar propagate through diffusion. A smooth power-law cutoff function describes the injection spectrum of the electrons/positrons.
\begin{equation}
\frac{d \rho }{d E} \propto E_k^{-\gamma_0}\left[1+\left(\frac{E_k}{E_b}\right)^{\frac{\gamma_1-\gamma_0}{s}}\right]^{-s}exp\left(-\frac{E_k}{E_{cut}}\right)
\end{equation}
Here, $\rm E_b$ is break energy, $\rm \gamma_0$, and $\rm \gamma_1$ are spectral indices for low and high energy bands, and $\rm s$ represents the smoothness parameter. In this work, we fixed $\rm E_b=10 ~GeV$, $\rm \gamma_0=1.0$, and $\rm s=0.5$.  We have adopted $\rm E_{cut} = 1.7 \times \sqrt{L_{36}} ~PeV \sim  300 ~TeV$ \citep{2023PhRvD.107l3020S}, where $\rm L_{36}$ represents the current spin-down luminosity of the Geminga pulsar in units of $\rm 10^{36}~erg s^{-1}$.

\section{Results} \label{sec:result}
In this section, we present our computational results, which include the multi-band non-thermal radiation energy spectrum of the Geminga halo, the distribution of diffuse gamma rays in the Galactic disk, and the map distribution of signal intensities for the background and halo in the Geminga region. We also discuss the mechanisms that lead to these distributions and shapes forming near the pulsar. In calculating the background signals of the diffuse gamma rays from the Galactic disk, the Geminga halo, and the spectral components of the background radiation in the halo region, it is necessary to compute the distribution of cosmic ray particles in the Milky Way. We use the B/C ratio and cosmic ray proton spectra observed near Earth to constrain the relevant parameters of the propagation model. \ref{appendix} shows the spatially dependent cosmic ray propagation model used to accurately reproduce the observed B/C and proton spectra, with the values of the relevant parameters listed in Table \ref{tab1}. 

\begin{figure}[t]
\centering
\includegraphics[width=0.95\linewidth]{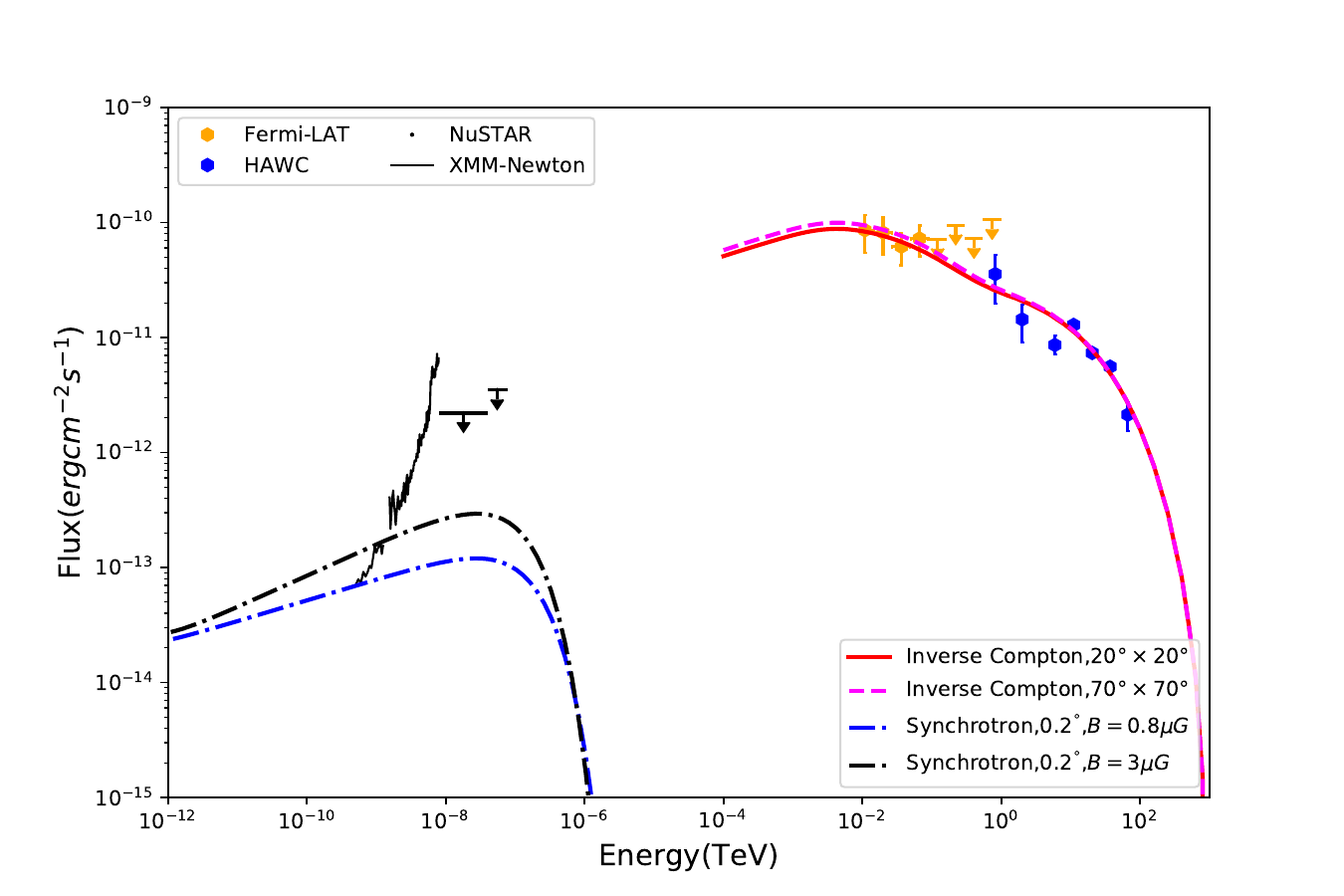}
\caption{The multi-wavelength non-thermal emission from the Geminga pulsar halo as compared to the upper limits in the X-ray band obtained using XMM–Newton and NuSTAR data \citep{2024A&A...689A.326M} and Fermi-LAT\citep{2019PhRvD.100l3015D} and HAWC \citep{2017Sci...358..911A} observations. The red line and magenta dashed line represent the angle-integrated inverse Compton radiative spectrum within the within a $20\degree\times20\degree$ and $70\degree\times70\degree$ around the Geminga, respectively. The black and blue($\rm B=3\mu G$ and $\rm B=0.8\mu G$) dot-dashed represents synchrotron emission within 0.2$\degree$, respectively.} 
\label{fig1}
\end{figure}

\begin{figure}[t]
\centering
\includegraphics[width=0.95\linewidth]{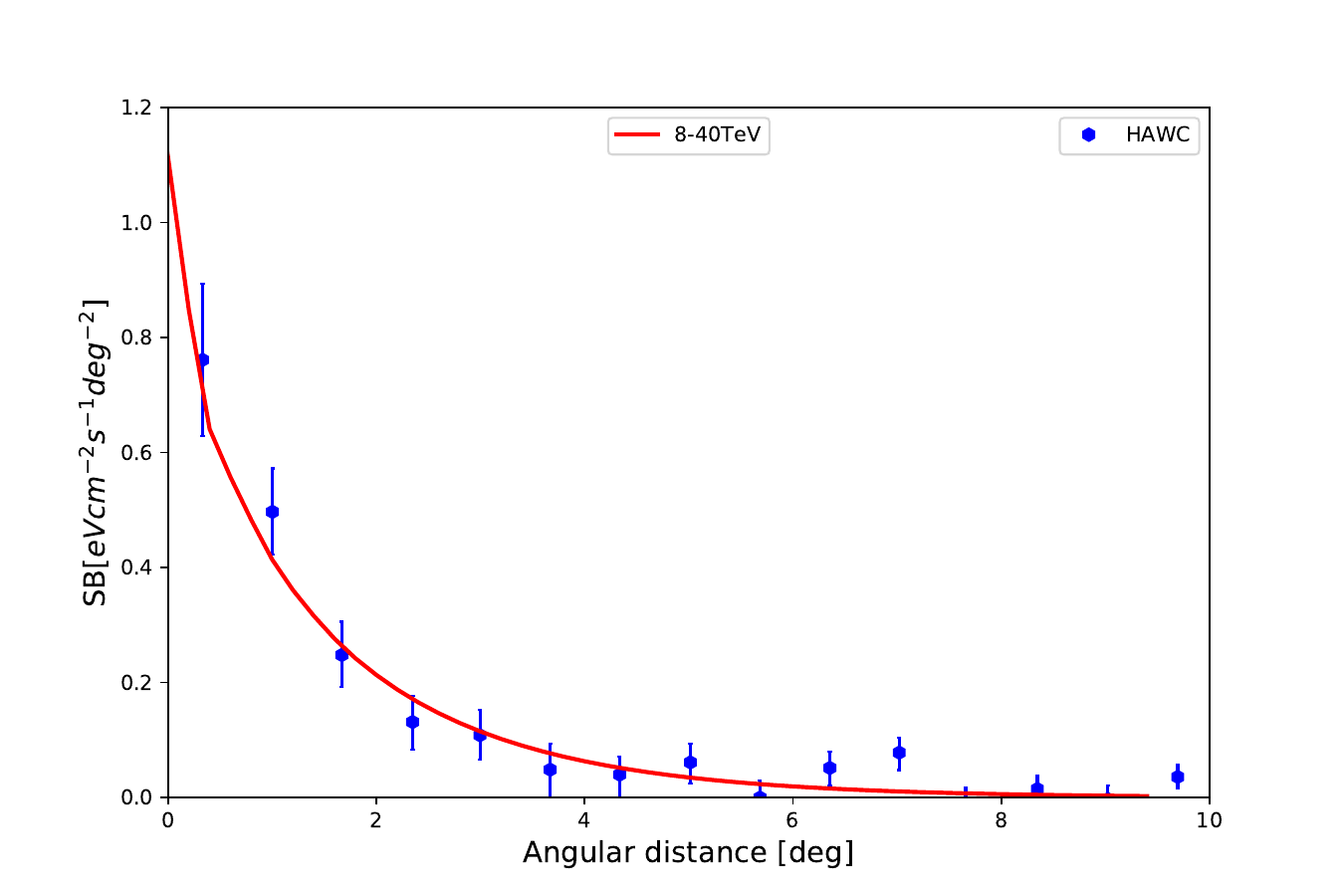}
\caption{Surface brightness of the IC emission shown as a function of angular distance around Geminga. We also display the HAWC data \citep{2017Sci...358..911A}.}
\label{SBP}
\end{figure}

\begin{figure*}[t]
\centering
\includegraphics[width=0.45\linewidth]{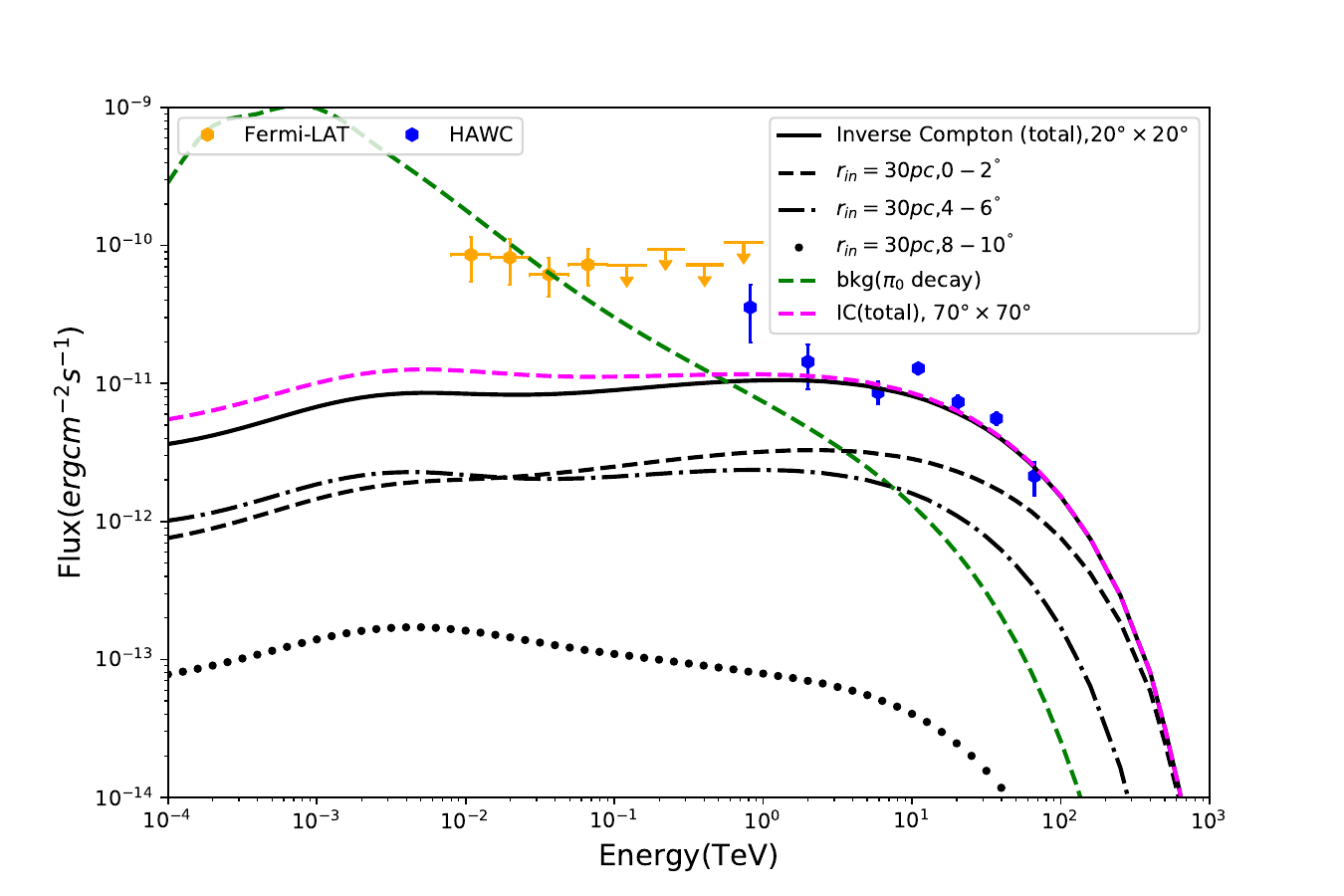}
\includegraphics[width=0.45\linewidth]{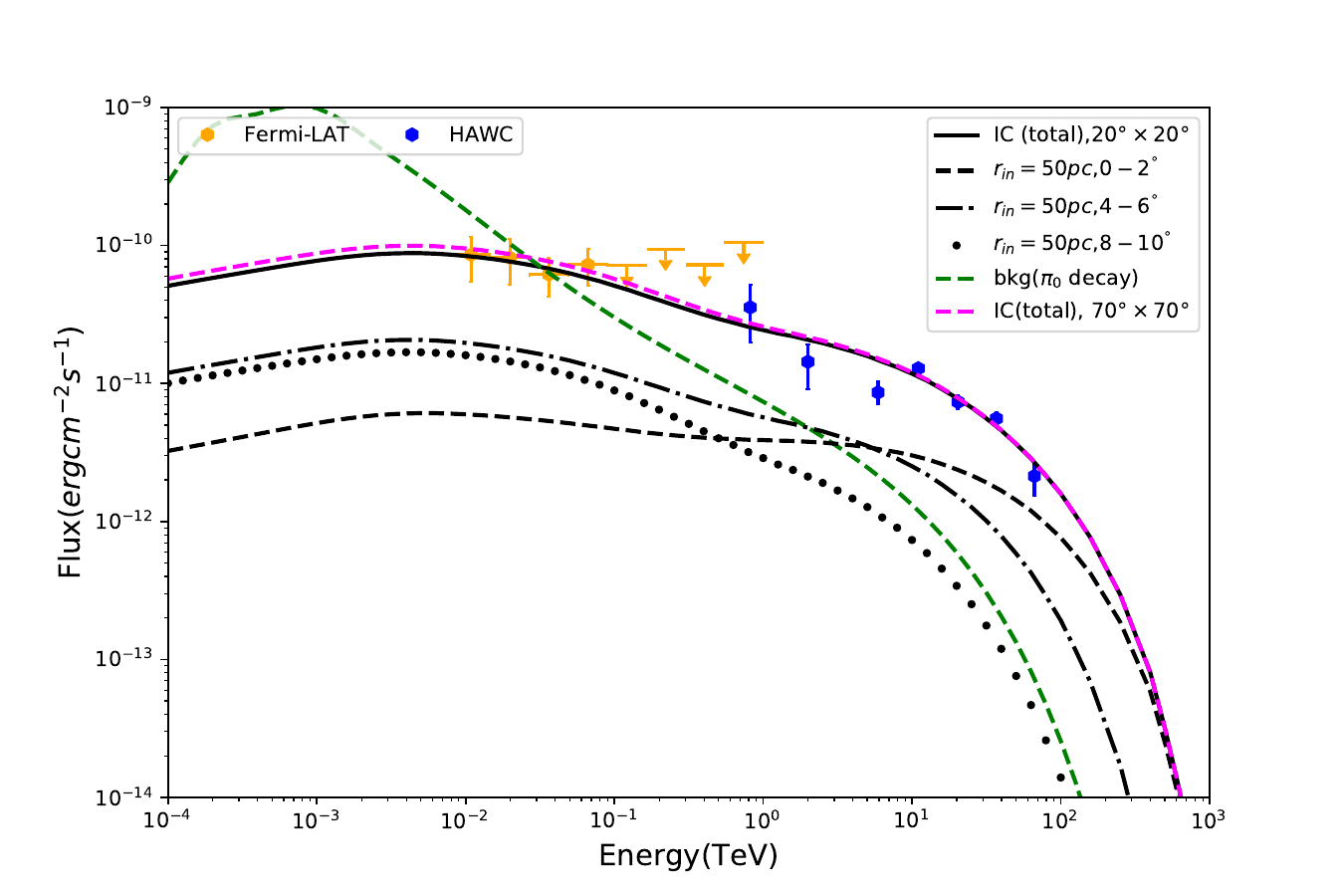}
\caption{The spectrum of IC emission in different ring regions is presented, with the left panel representing a slow-diffusion zone of about $\rm 30 pc$ in size, and the right panel corresponding to a zone of about $\rm 50 pc$. The black dashed, dash-dotted, and dotted lines represent the contributions from the 0-2 degrees, 4-6 degrees, and 8-10 degrees regions, respectively. The black solid line and magenta dashed line represent the total radiation within the within a $20\degree\times20\degree$ and $70\degree\times70\degree$ around Geminga, respectively. The green dashed line represents the background diffuse gamma-ray radiation component within this range.}
\label{fig1_2}
\end{figure*}

\begin{figure}[t]
    \centering
    \includegraphics[width=0.95\linewidth]{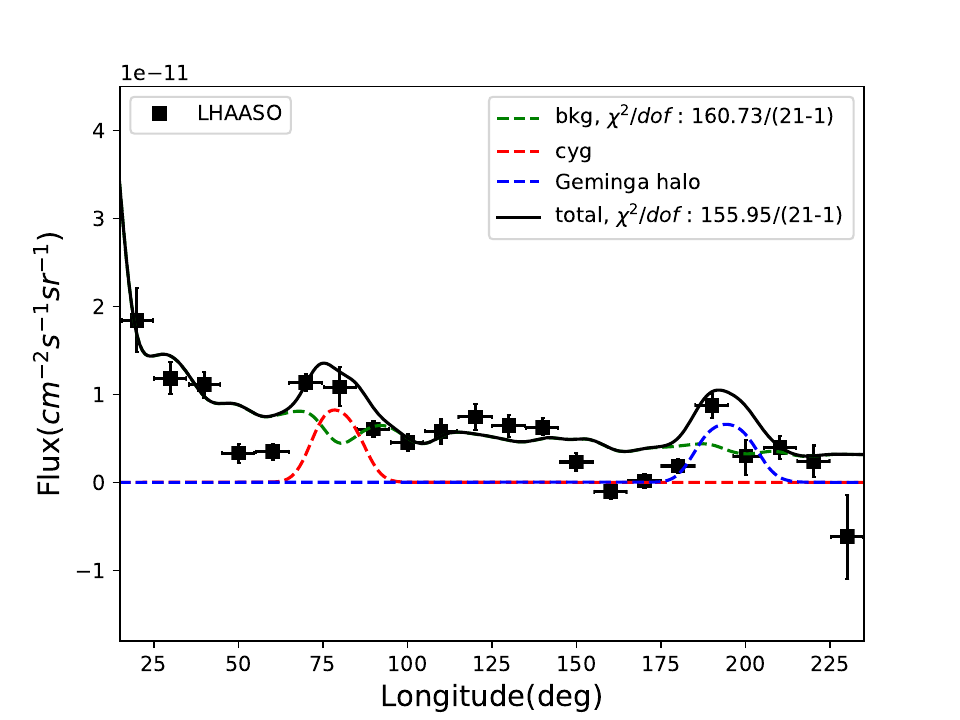}
    \caption{The flux of diffuse gamma-ray emissions in the Galactic longitude range of $\lvert$b$\rvert$$<$5$\degree$ with the energy range 10-63 TeV, in comparison with the LHAASO observations\citep{2023PhRvL.131o1001C}. The red dashed line represents the diffuse gamma components from the CR protons produced by the Cygnus bubble, the green dashed line represents the emission flux of diffuse gamma predicted by the CR SDP model, and the blue dashed line is components from the Geminga halo.}
    \label{fig2}
    \end{figure}

\begin{figure*}[htbp]
\centering
\includegraphics[width=0.48\linewidth]{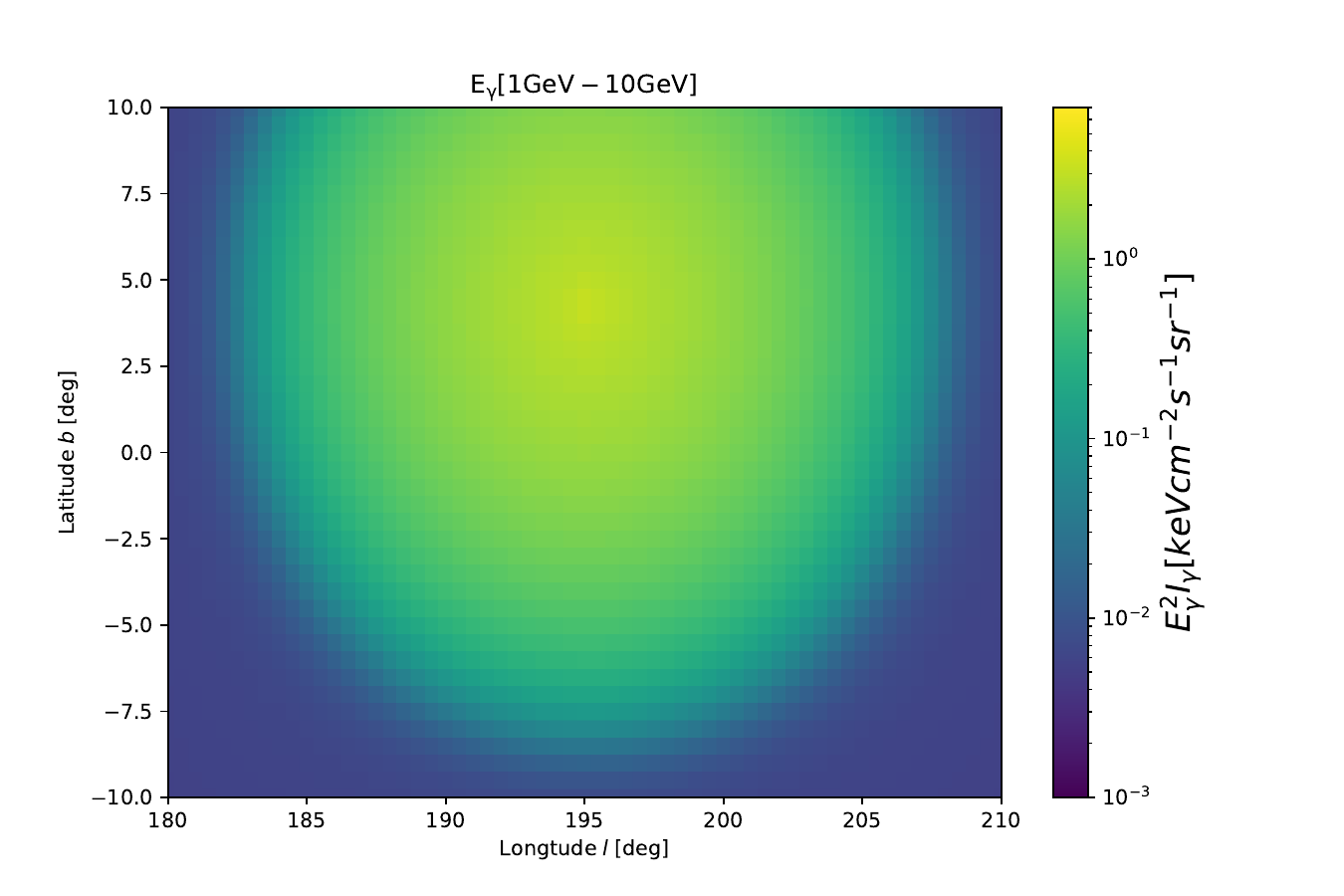}
\includegraphics[width=0.48\linewidth]{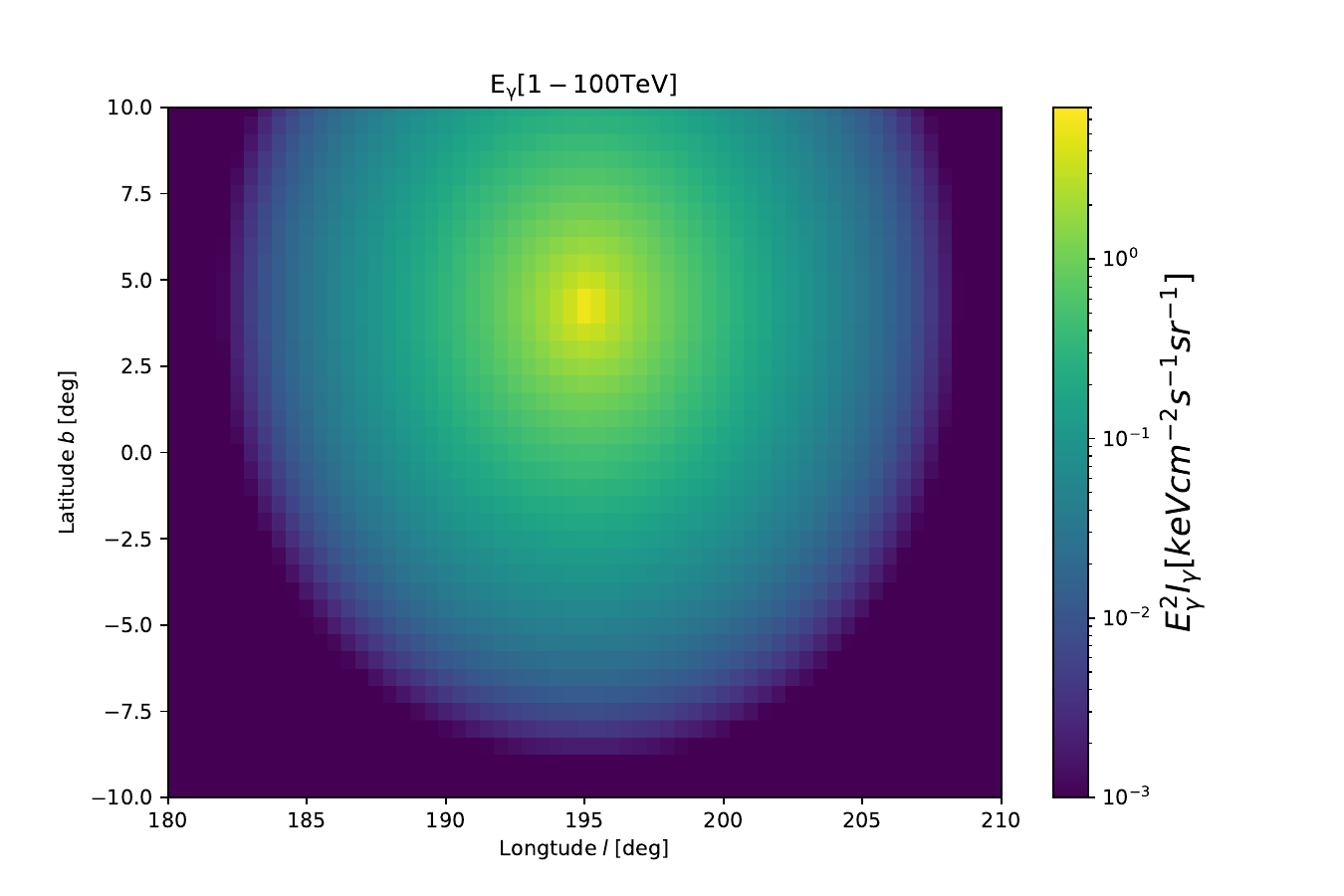}
\includegraphics[width=0.48\linewidth]{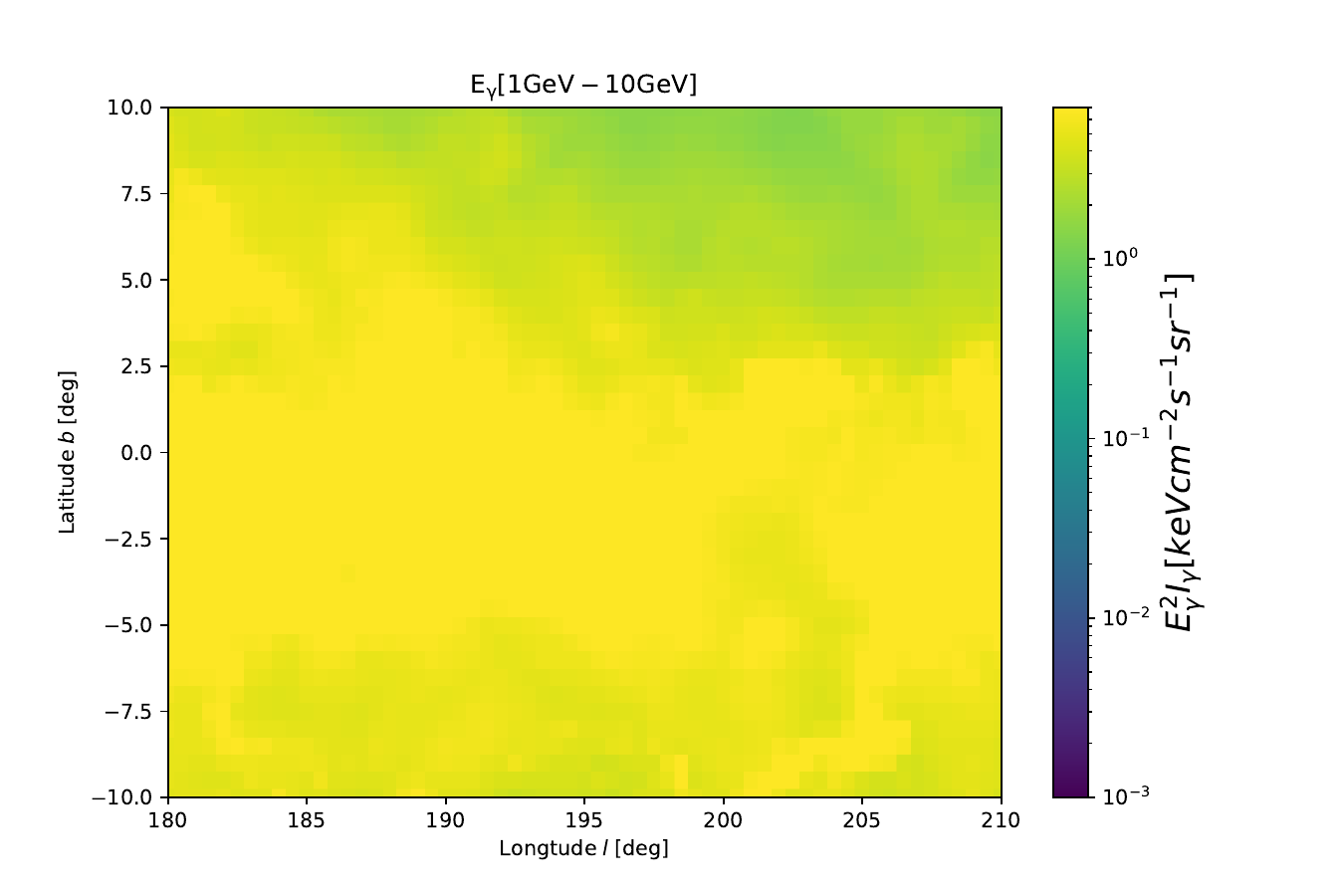}
\includegraphics[width=0.48\linewidth]{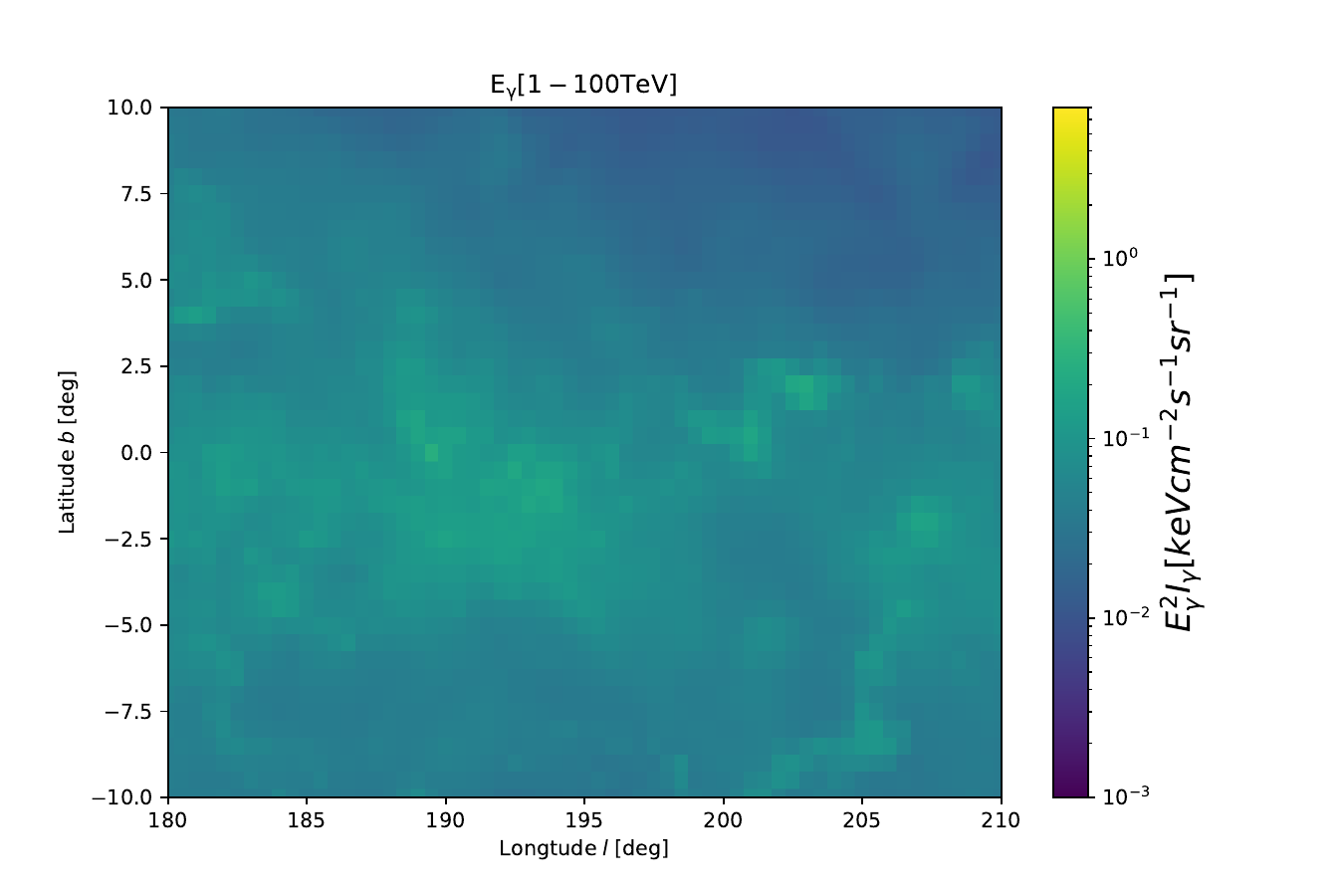}
\includegraphics[width=0.48\linewidth]{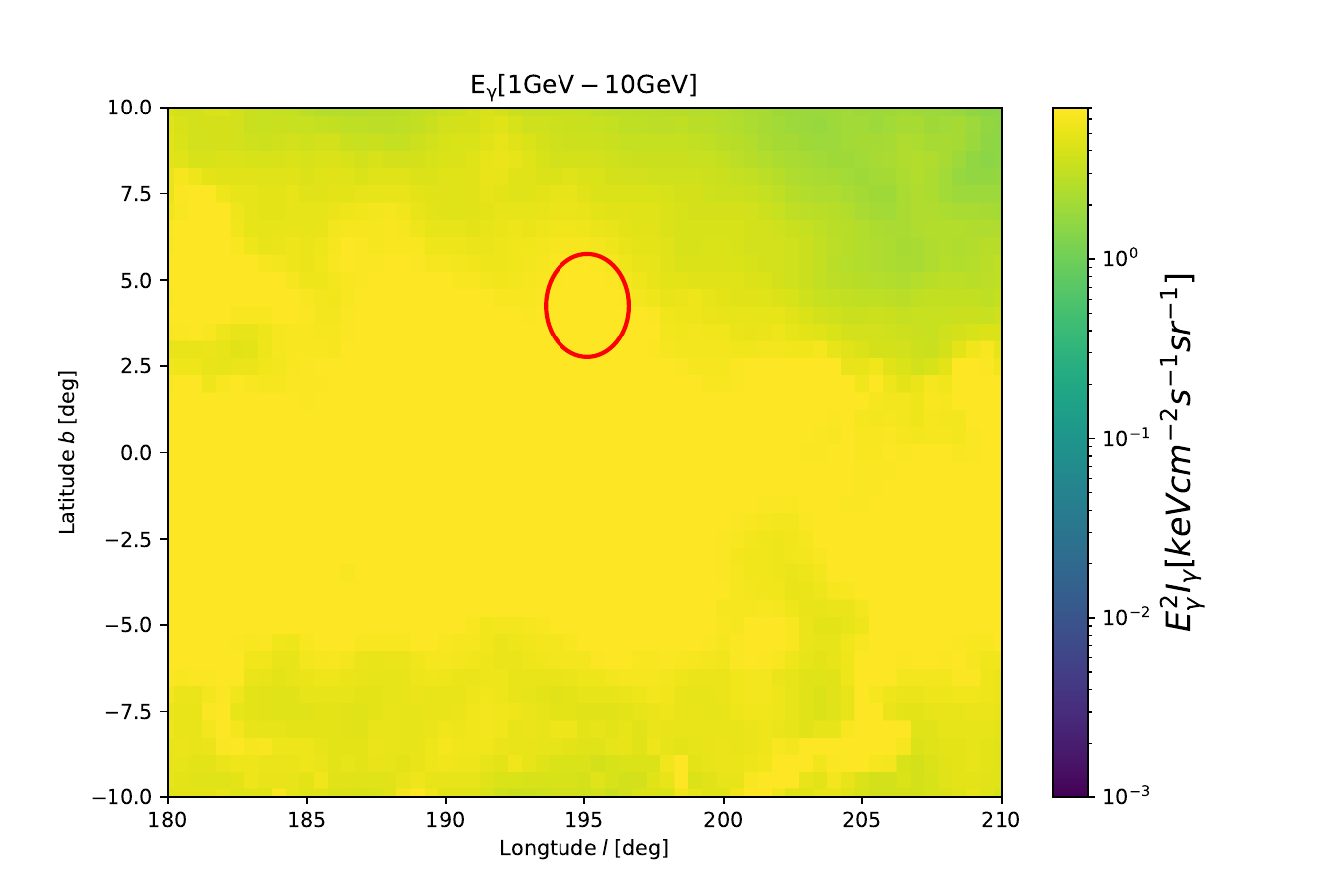}
\includegraphics[width=0.48\linewidth]{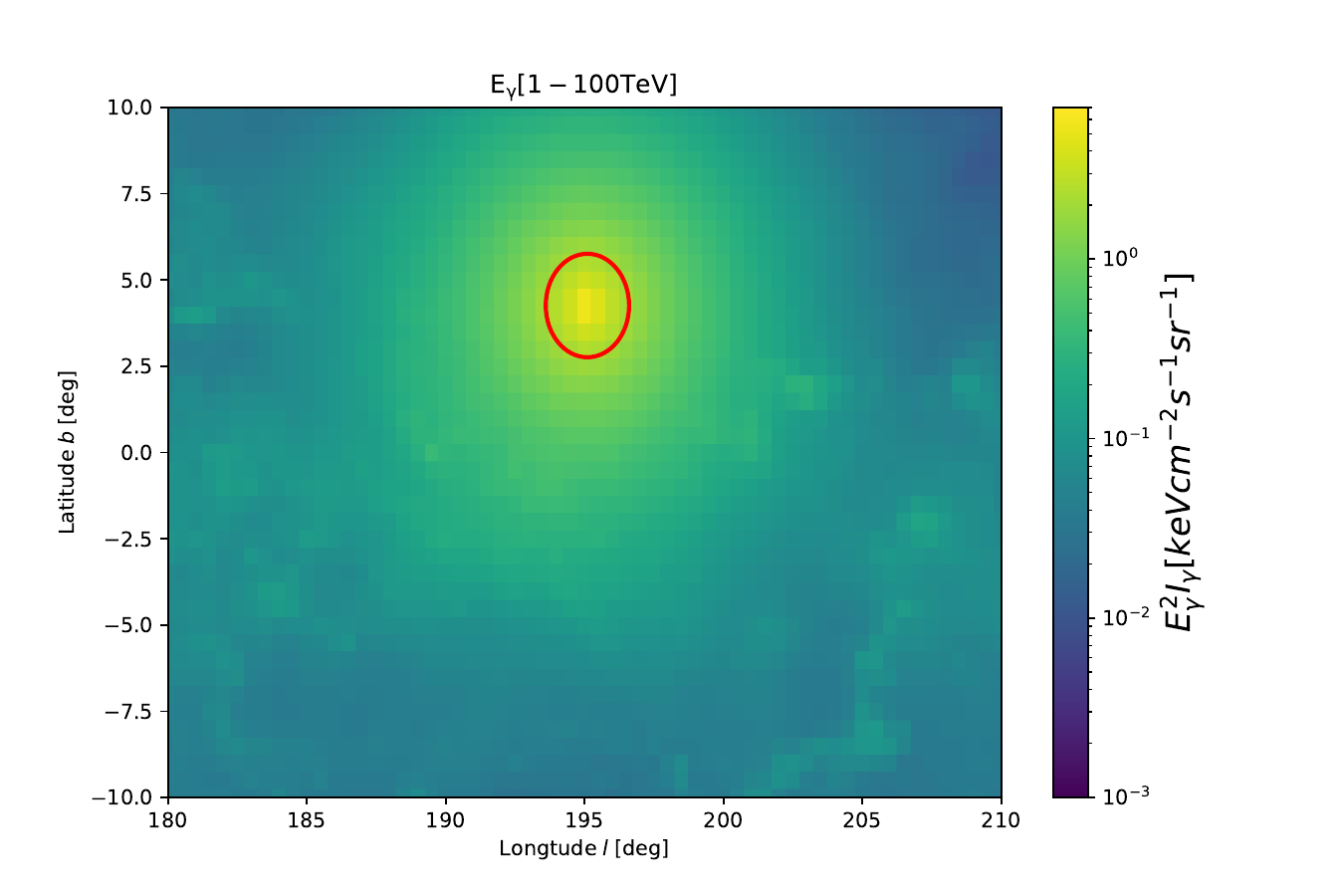}
\caption{Gamma-ray flux maps (left panel), which include individual (top left panel: signal from Geminga pulsar halo, and middle left panel: background components) and Comprehensive (bottom left panel: total emission components) maps in the energy band 1–10 GeV. Right panel maps are the same as left panel but in the range 1–100 TeV.}
\label{fig3}
\end{figure*}

\begin{figure*}[htbp]
\centering
\includegraphics[width=0.48\linewidth]{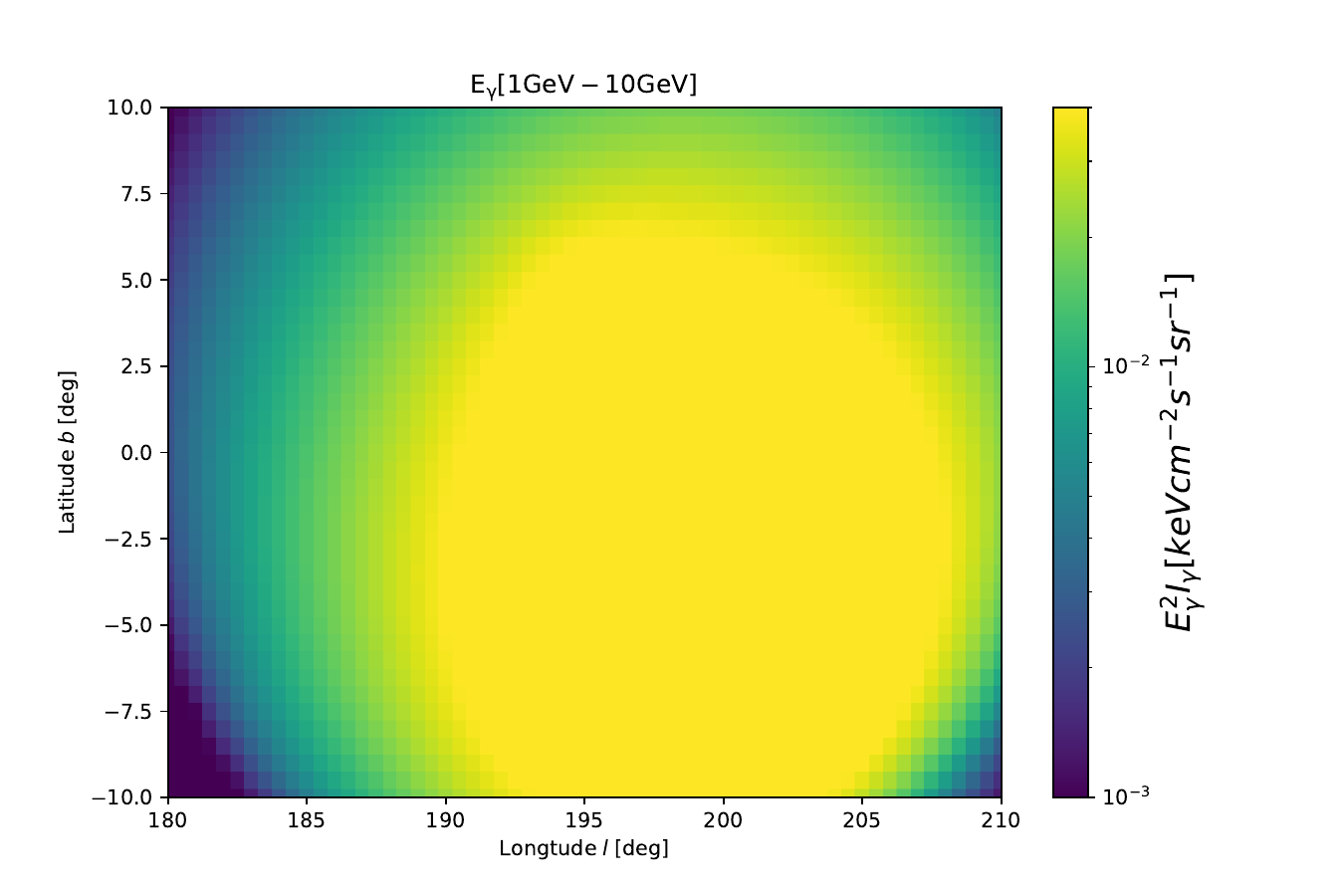}
\includegraphics[width=0.48\linewidth]{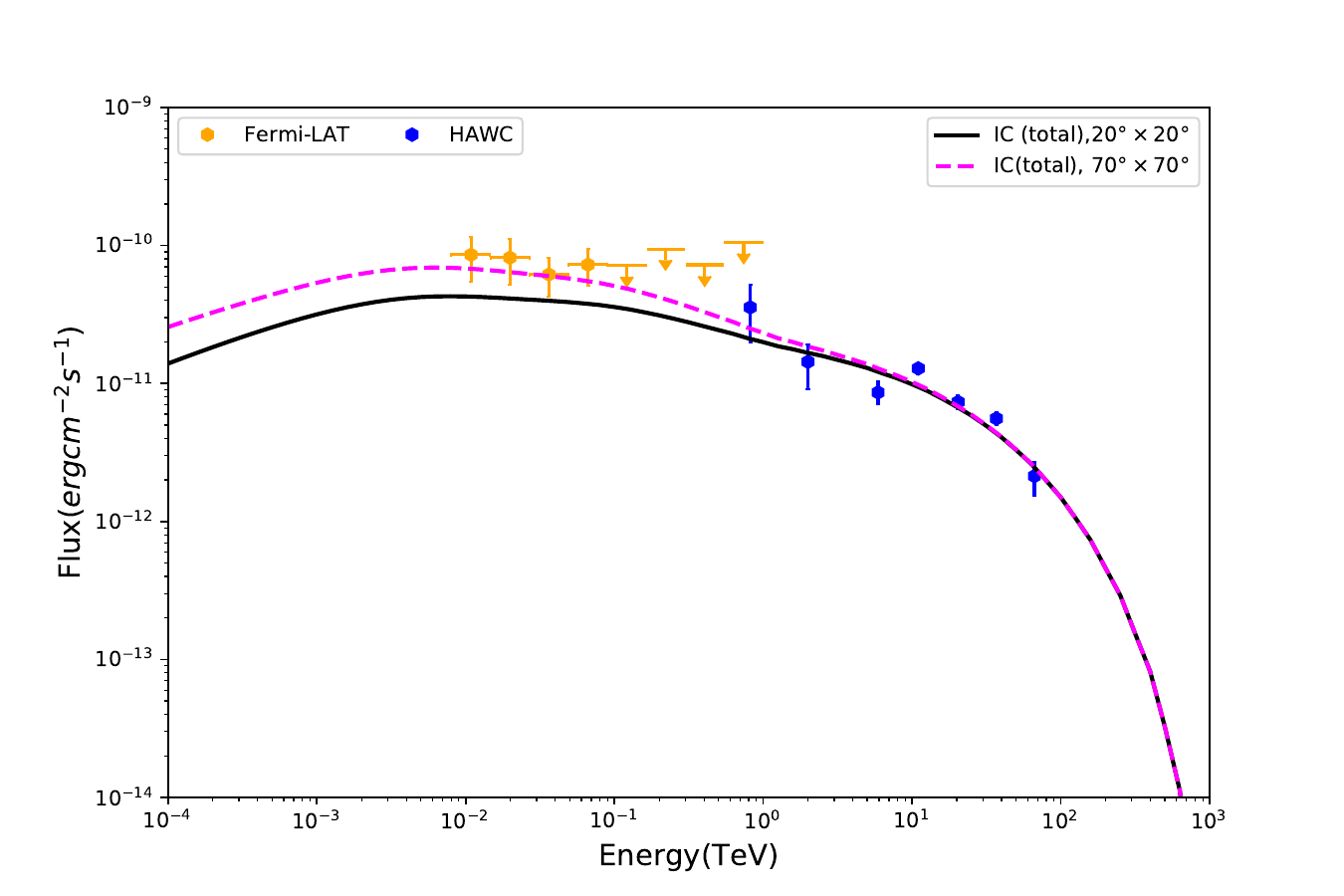}
\includegraphics[width=0.48\linewidth]{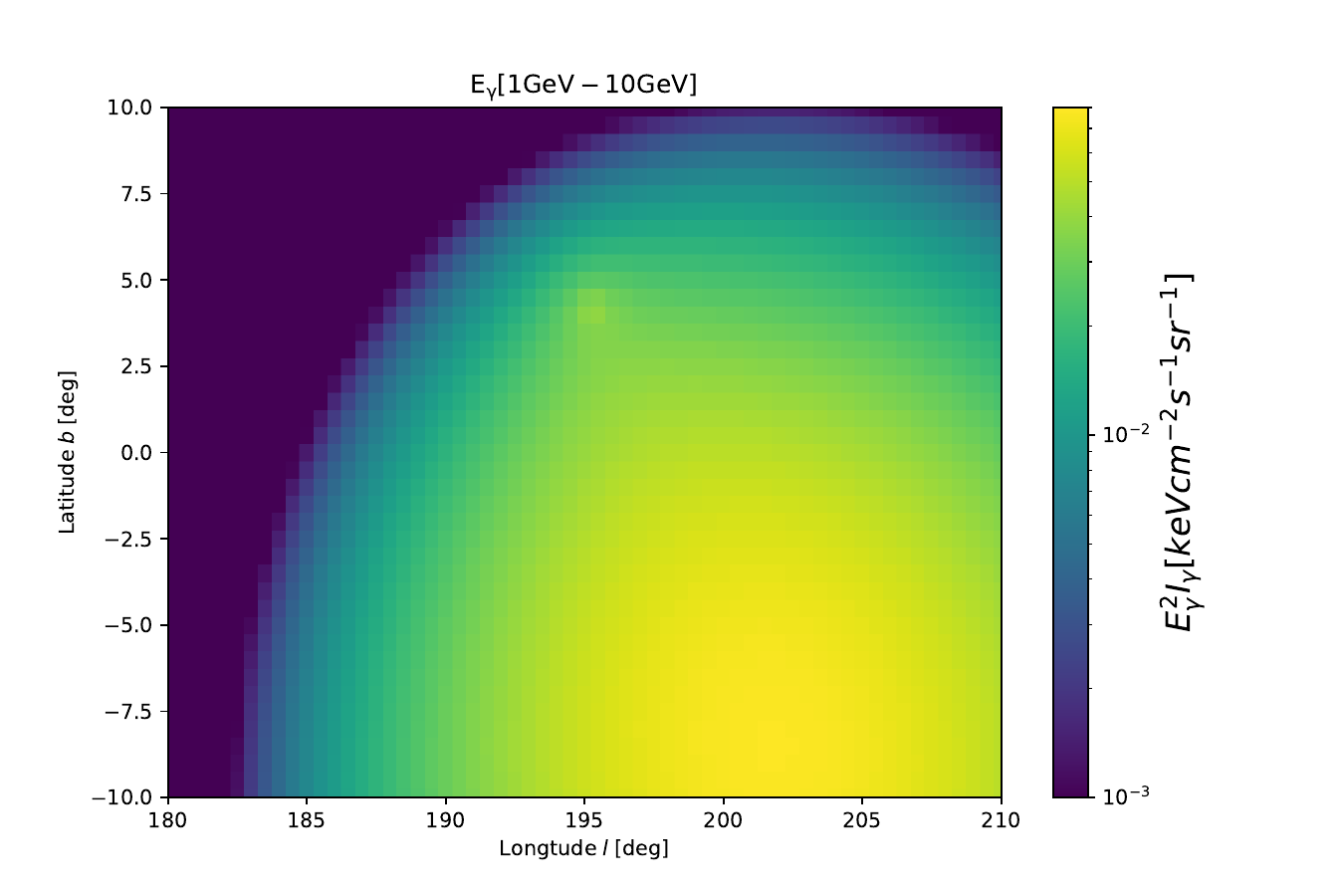}
\includegraphics[width=0.48\linewidth]{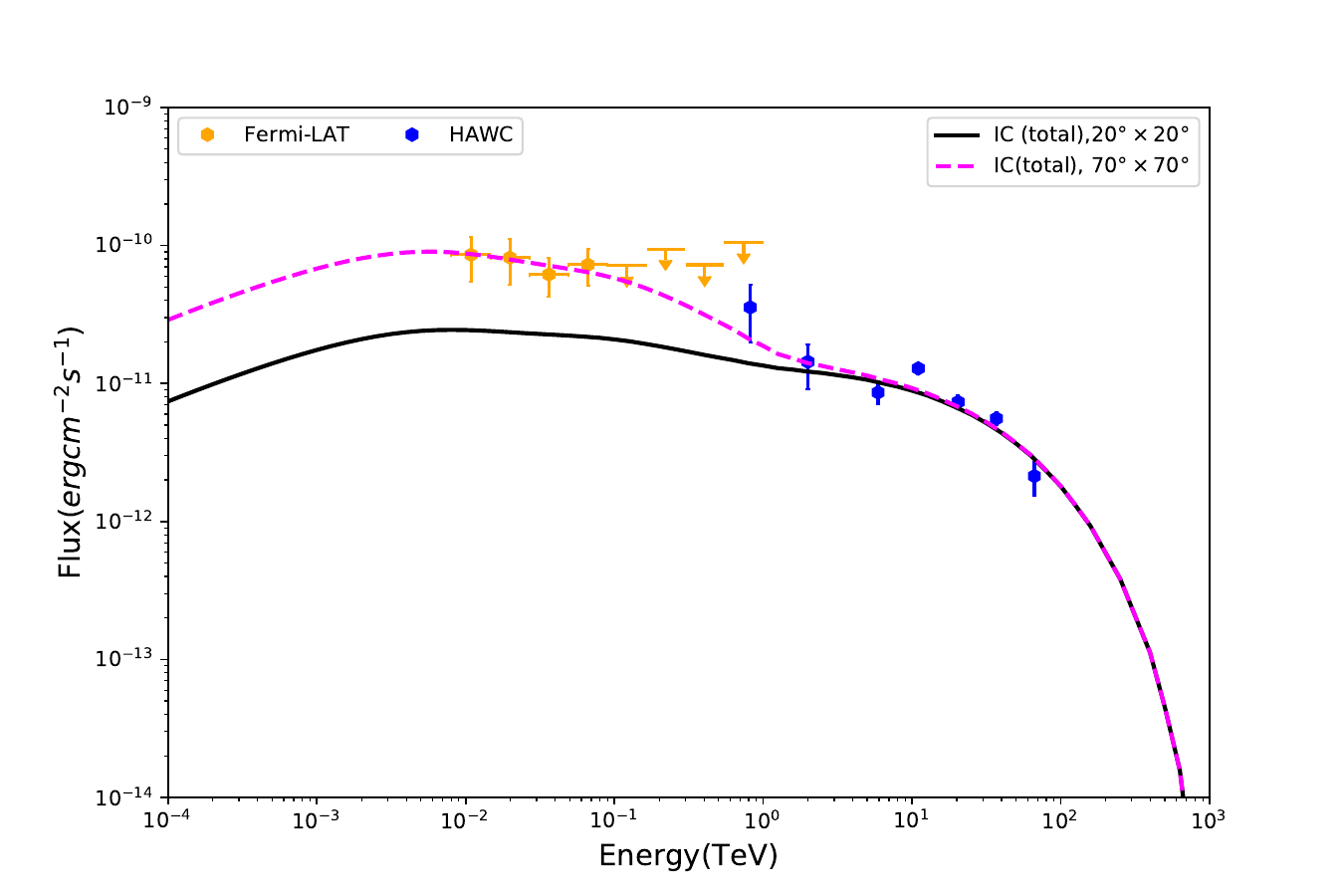}
\caption{Top panel (co-moving SDZ):  the center of the Slow Diffusion Zone (SDZ) coincides with the
Geminga pulsar, and the SDZ moves with the pulsar. Bottom panel (static SDZ): the SDZ center remains at the pulsar’s birth location without co-motion. The left panels show the emission morphology of the Geminga halo alone, while the right panels display the corresponding energy spectrum.}
\label{motion}
\end{figure*}

\begin{figure}[t]
    \centering
    \includegraphics[width=0.95\linewidth]{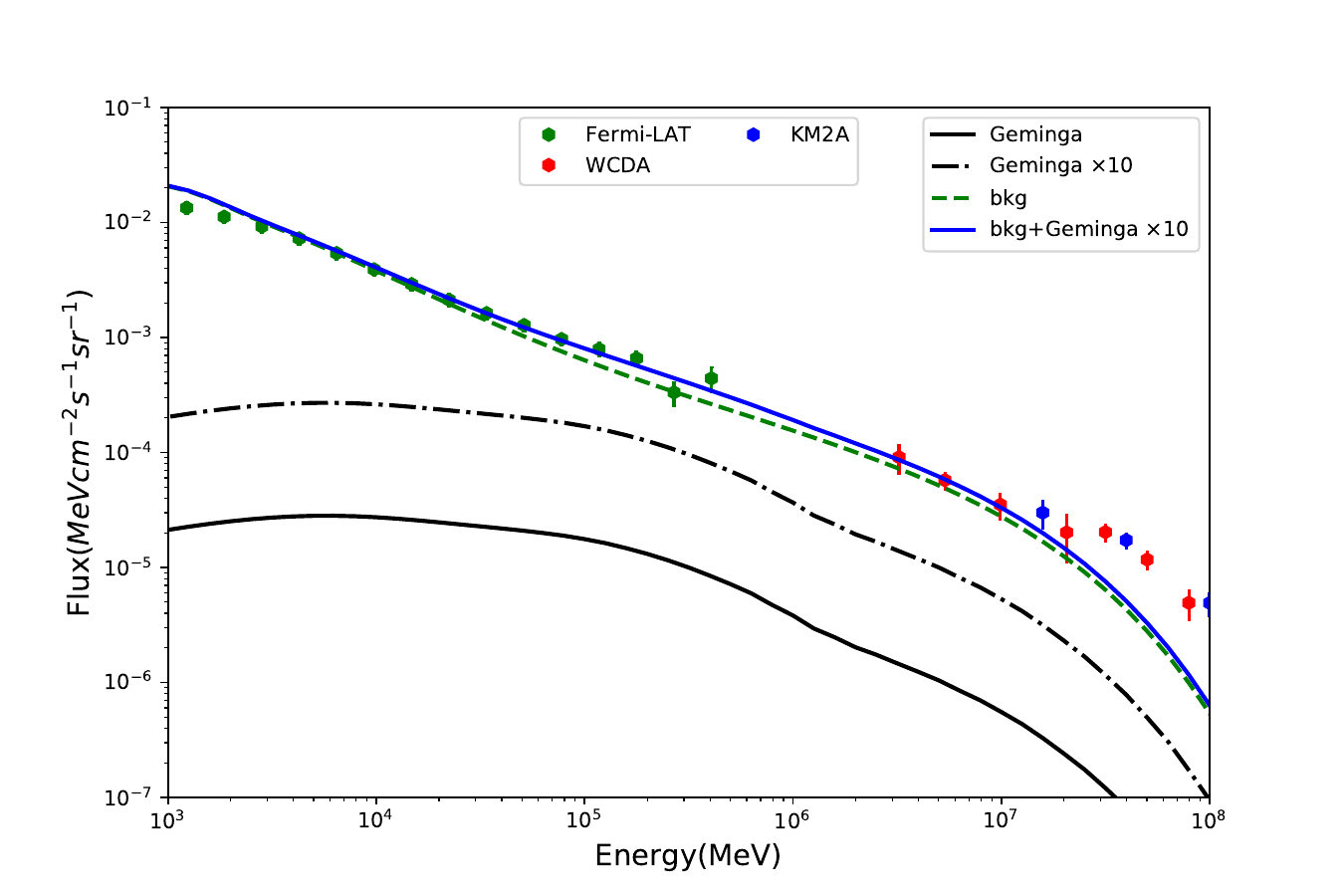}
    \caption{ SEDs of the diffuse $\gamma$-ray emission in the outer regions of the Galaxy ($125\degree<l<225\degree$, $|b|<5\degree$).  The data points are taken from the Fermi-LAT \citep{2023ApJ...957...43Z}, KM2A \citep{2023PhRvL.131o1001C} and WCDA \citep{2025PhRvL.134h1002C}.}
    \label{diffuse_gamma}
    \end{figure}
    
\begin{figure}[t]
    \centering
    \includegraphics[width=0.95\linewidth]{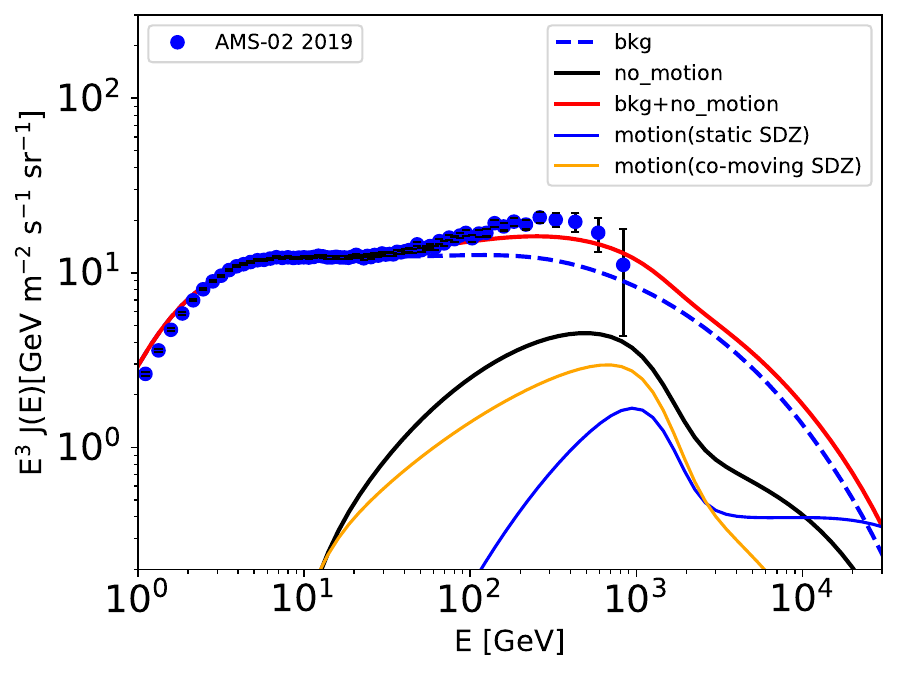}
    \caption{The predicted spectrum of positrons and observed data from AMS-02 \citep{2019PhRvL.122d1102A}. The blue dashed line represents the background cosmic-ray positron component. The black solid, orange solid, and blue solid lines show the contributions from Geminga for three scenarios: without pulsar proper motion, with proper motion in a static slow-diffusion zone (SDZ), and with proper motion in a co-moving SDZ, respectively.} 
    \label{positron}
    \end{figure}

\subsection{Emissions from the Geminga pulsar halo}
To reproduce the observed phenomena from the Geminga halo, we use the model constructed using the method described in Section \ref{sec:method} to calculate the radiation energy spectrum and its morphological distribution from the Geminga halo.

Figure \ref{fig1} and \ref{SBP} present the multi-wavelength non-thermal radiation and the angular profile of the surface brightness from the Geminga halo, respectively. Following the observational data, the synchrotron component is limited to the contribution within 0.2 degrees around the pulsar. In contrast, the high-energy IC component comes from the contribution of electron/positron pairs interacting with the cosmic background radiation via inverse Compton scattering in the 10$\degree$ region. By using the upper limit data from the X-ray upper limits (XMM–Newton, NuSTAR) and the observational data from Fermi-LAT and HAWC in the high-energy range, the magnetic field of the halo was required to be $\rm B_0 < 0.8 ~\mu G$, which is significantly weaker than the typical magnetic field $\rm B=3\mu G$ in the interstellar medium. As shown in Figure \ref{fig1}, It is evident that the theoretical prediction for $\rm B=3~\mu G$ does not align with the XMM-Newton upper limits, suggesting it may be less effective in constraining the model parameters compared to utilizing the SED data. The other constrained quantity, the scale of the pulsar halo's slow diffusion region is about 50 pc, with a diffusion coefficient $\rm D_h=1.98 \times 10^{26}~cm^2s^{-1}$.

Figure \ref{fig1_2} shows the emissions produced by the electrons and positrons within the different extended regions around the Geminga pulsar. It can be concluded that low-energy radiation primarily originates from the extended regions farther away from the pulsar, while the high-energy component is dominated by contributions from electrons in the regions closer to the pulsar.
This energy-dependent scenario arises because cooling effect of cosmic ray electrons, low-energy electrons have a longer propagation distance compared to high-energy electrons.

In Figure \ref{fig2}, we present the distribution of the diffuse gamma rays in the Galactic disk. The background component of Figure \ref{fig2} represents the radial distribution of diffuse gamma-ray radiation produced by the interaction of background cosmic rays with ISM gas in the Galactic disk($|b|<5^{\degree}$). We calculate the spatial distribution of cosmic rays using the spatially dependent propagation model described in the \ref{appendix} and utilize the GALPROP package\citep{1998ApJ...509..212S,2022ApJS..262...30P} to compute the Galactic diffuse gamma-ray radiation, further simulating the radial distribution of diffuse gamma-ray radiation in the Galactic disk. On the other hand, to compare with the LHAASO observational data, we process the model-calculated data like the LHAASO data: 1) smoothing the theoretical data and 2) subtracting the contributions from point sources and extended sources when calculating the background radiation. Therefore, in this work, the sources detected by KM2A and those observed by other experiments summarized in TeVCat are all subtracted during the calculation of the diffuse radiation, especially in the regions of the Cygnus bubble and Geminga. The region of the Cygnus bubble is masked within about 6$\degree$, while the region around Geminga had its contribution removed within an 8$\degree$ radius, which is about 2.5 times larger than the masked regions of other sources listed in TeVCat. As shown in Figure \ref{fig2}, even with the removal of contributions from extended sources in a larger area, the Cygnus bubble and Geminga halo still provide dominant contributions. The detailed calculation process of the Cygnus bubble component in Figure \ref{fig2} can be found in our previous work \citep{2024ApJ...974..276N}. Similar to the diffusion of electrons/positrons in the halo in this work, the Cygnus bubble is also considered a slow diffusion region, and its energy spectrum, morphological distribution, and diffuse gamma-ray structure in the region have been well reproduced.

\subsection{Map of Geminga halo}
In the vicinity of the Geminga pulsar, no radiation morphology was observed in the 1-8 GeV energy range, but there is extended radiation in the TeV range. To investigate the mechanism behind this phenomenon, we use the observational data of the multi-wavelength energy spectrum from the Geminga halo, as well as the model parameters derived from the constraints of the diffuse gamma-ray observations of the Galactic disk by LHAASO, to calculate the map signals of the Geminga halo and the background. As shown in Figure \ref{fig3}, in the 1-10 GeV energy range, the Geminga halo signal is relatively weak compared to the background intensity, Consequently, under background contamination, the extended emission morphology in the vicinity of the Geminga pulsar exhibits a low significance level.
However, in the 1-100 TeV energy range, Geminga produces a signal much stronger than the background, and a significant radiation morphology remains even with the background contribution considered, consistent with the observations from the HAWC. 

Theoretical models suggest that lower-energy electrons/positrons produce the GeV radiation through inverse Compton scattering with soft background photons. However, once the electrons/positrons escape from the pulsar, they propagate through the slow diffusion region near it and diffuse into the more distant and homogeneous normal Galactic environment. In the region where ( $r<r_i$), the diffusion coefficient of the particles is at least two orders of magnitude smaller than in other regions due to the spatial-dependence of $D_h$, meaning that the density of electrons/positrons that generate GeV radiation in this region is high. This situation appears inconsistent with the absence of significant extended emission detected within the $40\degree \times 40\degree $ region centered on Geminga \citep{2019ApJ...878..104X}.
On the other hand, it has been suggested that the proper motion of Geminga becomes significant at GeV energies \citep{2019PhRvD.100l3015D}. Therefore, the nondetection of few GeV emission can be explained by the scenario in which the region of interest ( $40\degree \times 40\degree $) considered is not large enough to detect the Geminga ICS halo due to the effect of Geminga's proper motion.

we have incorporated and discussed the effects of the pulsar's proper motion. The measured proper motion of the Geminga pulsar is $-80.0~mas~yr^{-1}$ in longitude and $156.0 ~mas~yr^{-1}$ in latitude \citep{2007Ap&SS.308..225F}. Assuming the current velocity is perpendicular to the line of sight, the pulsar's spatial velocity vector is $\rm (v_x,v_y,v_z) = (24.3,-89.9,182.6)~km~s^{-1}$.
We consider two scenarios:
Firstly, the center of the Slow Diffusion Zone (SDZ) coincides with the Geminga pulsar, and the SDZ moves with the pulsar while its size scales proportionally with the square root of time.
Secondly, the SDZ center remains at the pulsar's birth location without co-motion.
Normalization conditions are set as follows: 
For the first scenario, the SDZ size at the evolutionary endpoint (present time) is $\rm (r_i, r_o) = (50, 70)~pc$. For the second scenario, $\rm (r_i, r_o) = (90, 110)~pc$. The results are presented in Figure \ref{motion}. It shows that for both scenarios (co-moving and static SDZ), pulsar proper motion significantly affects GeV-band radiation. 

On the other hand, the diffuse gamma-ray ``excess” has been observationally confirmed \citep{2007APh....27...10P,2015ApJ...806...20B,2021PhRvL.126n1101A}. We believe that signal leakage from sources like Geminga may provide a significant contribution to this phenomenon.
Both scenarios demonstrate significant consistency with existing observational evidence: Fermi-LAT detected no significant 10–100 GeV emission within the $40\degree\times40\degree$ region surrounding Geminga \citep{2019ApJ...878..104X}, while significant radiation is observed in the $70\degree\times70\degree$ field at these energies \citep{2019PhRvD.100l3015D}. We therefore calculate the contribution of Geminga halo signal leakage to the diffuse gamma-ray ``excess" on the Galactic plane under this framework, as shown in the figure \ref{diffuse_gamma}.
As illustrated in the figure \ref{diffuse_gamma}, multiplying Geminga's leaked signal by a factor of 10 provides an excellent fit to the observed diffuse gamma-ray data. Here, we interpret this factor as indicative of multiple sources with similar leakage characteristics to Geminga distributed throughout this region.



\subsection{Contribution of the Geminga to the positron spectrum}
The excess of cosmic ray positrons is observed in the experimental measurements of the cosmic ray positron flux in the energy range of $\rm 10-500 ~GeV$ \citep{2019PhRvL.122d1102A,2013PhRvL.110n1102A,2009Natur.458..607A}, which is higher than the expected flux according to the ``standard cosmic ray propagation model". 
However, our calculations show that while the spatial-dependent cosmic ray propagation model predicts a larger contribution at high energies (as illustrated by the blue dashed line in Figure \ref{positron}), it still fails to match the observed experimental magnitude. 
Nevertheless, estimates based on electron energy losses suggest that sources on scales of $\rm \sim kpc$ might significantly contribute to the positrons observed on Earth \citep{1995A&A...294L..41A}. 

Geminga is a middle-aged pulsar located about $\rm 250~pc$ from Earth, with an age of $\rm \sim 340~ky$ and a spin-down power of $\rm 3.2\times 10^{34}~ergs^{-1}$ \citep{2007Ap&SS.308..225F}. Its distance and age provide strong grounds to believe it could be a major source of the positron excess. The spectrum of positrons obtained near Earth, as shown in Figure \ref{positron}, is constrained by the data from Fermi, HAWC, and LHAASO. 
While the spectrum is nearly consistent with the observational data, it is difficult to distinguish between electrons and positrons in the halo, so Geminga may contribute less than $50\%$ of the observed positrons. Besides, the Geminga contributes to the positron flux at Earth strongly depend on the chosen injection slope of the spectrum $\gamma_1$ and size of SDZ\citep{2019PhRvD.100l3015D,2019ApJ...879...91J}. The softer injection spectra have more low-energy electrons, and there is more confinement for low-energy electrons for larger sizes of the SDZ. A smaller SDZ leads to a larger flux of positrons at Earth because of the larger effective diffusion coefficient, and larger values of $\gamma_1$ also result in a larger positron flux at Earth in the observed energy range.
In addition, it may be suggested that other nearby candidate sources or alternative physical processes, such as dark matter annihilation or proton-proton (p-p) interactions of cosmic rays from nearby sources, could also play a significant role and should not be ignored.

\section{Conclusion and Summary} \label{sec:conclusion}
This paper investigates the transport and radiation properties of electrons/positrons in the Geminga halo using observational data from LHAASO, Fermi-LAT, and HAWC, with the GALPROP package as the simulation tool. It expands on the role of pulsar halos in the cosmic ray propagation process, explores the spatial dependence of diffuse gamma radiation, and tests the physical framework of the spatially dependent cosmic ray propagation. 

The previous findings suggest that Fermi-LAT observations reveal no significant extended emission within the $40\degree\times40\degree$ region surrounding Geminga in the energy range of 10 to 100 GeV \citep{2019ApJ...878..104X}, while the corresponding emission is detected in a larger $70\degree\times70\degree$ field \citep{2019PhRvD.100l3015D}. We reproduce the non-thermal radiation energy spectrum from the Geminga halo, and simulate the extended radiation maps in the 1-10 GeV and 1-100 TeV energy ranges. Our calculated results support that this may result from the pulsar's proper motion combined with its birth in a slow-diffusion region.

We find that the signal leakage from the local Geminga halo may contribute to some extent to the fluctuating structure of the diffuse gamma-ray radiation observed by LHAASO. 
Furthermore, such leakage from local sources may contribute to the diffuse gamma-ray ``excess". However, the contributions from signal leakage and local gas uncertainties cannot be disentangled, leaving it uncertain which effect dominates the measured fluctuations. 
We hope that LHAASO will be able to observe the extended radiation from the Geminga halo and other TeV halos in the future, and more precise observations will further study and disentangle signal leakage and other effects. 

\section*{Acknowledgements}
We thank  Silvia Manconi for providing the XXM data. This work is supported in China by National Key R$\&$D program of China under the grant 2024YFA1611402, and supported by the National Natural Science Foundation of China (12333006, 12275279, 12375103).

\begin{appendices}  
\renewcommand{\thesection}{Appendix \Alph{section}} 
\section{Spatially-dependent Propagation Model}\label{appendix}
Cosmic rays are accelerated during the active evolution phases of cosmic ray sources (such as supernova remnants \citep{2023ApJ...952..100N,2013Sci...339..807A,2014IJMPD..2330013A}, pulsar wind nebulae \citep{2022ApJ...924...42N,2021Natur.594...33C,2021Sci...373..425L,2015MNRAS.451.3145Z}, the galactic center \citep{2021NatAs...5..465A}, binary star systems \citep{2012Sci...335..175M} and so on) and injected into the interstellar medium, where they freely diffuse. The diffusion timescale can last up to approximately 10 million years \citep{1977ApJ...217..859G}. During this diffusion process, cosmic rays interact with the surrounding material, producing secondary cosmic ray particles. For example, the fragmentation and decay of heavy nuclei produce secondary cosmic ray particles, including light nuclei (Li, Be, B) and antiparticles (antiprotons and positrons). Additionally, cosmic ray electrons and positrons interact with the ambient magnetic field and background radiation throughout their journey, resulting in energy loss. The entire dynamical behavior of cosmic ray propagation can be described by the following diffusion equation \citep{2007ARNPS..57..285S,2017JCAP...02..015E}:

\begin{figure}[t]
\centering
\includegraphics[width=0.95\linewidth]{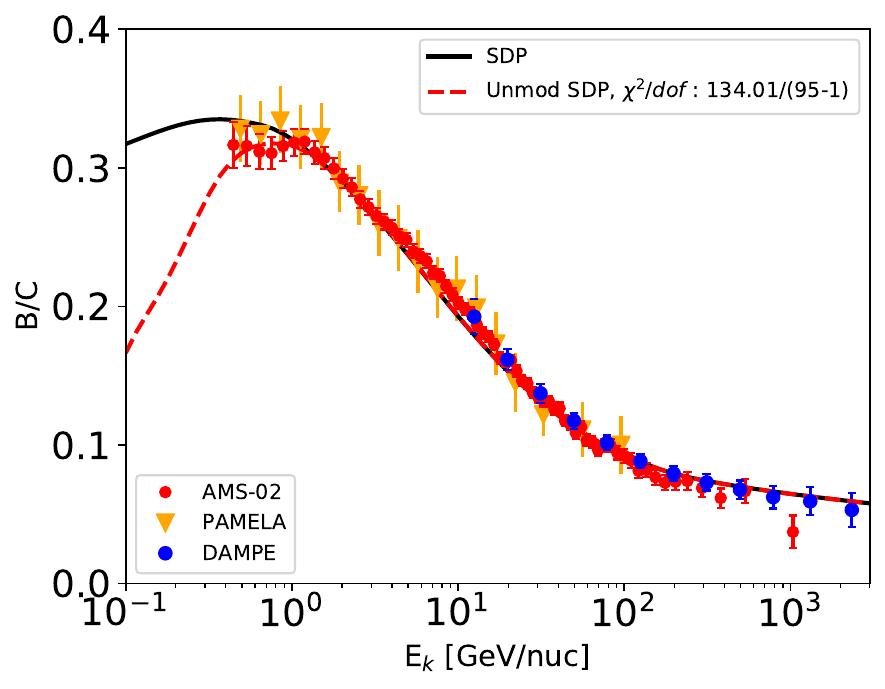}
\caption{A comparison of the B/C ratio calculated using the CR SDP model with observational data from AMS-02 \citep{2017PhRvL.119y1101A}, PAMELA \citep{2014ApJ...791...93A}, and DAMPE \citep{2022SciBu..67.2162D}. The red dashed line indicates the spectrum calculated without considering solar modulation. In this study, the solar modulation potential is consistently assumed to be 550 MeV.}
\label{fig4}
\end{figure}

\begin{figure}[t]
    \centering
    \includegraphics[width=0.95\linewidth]{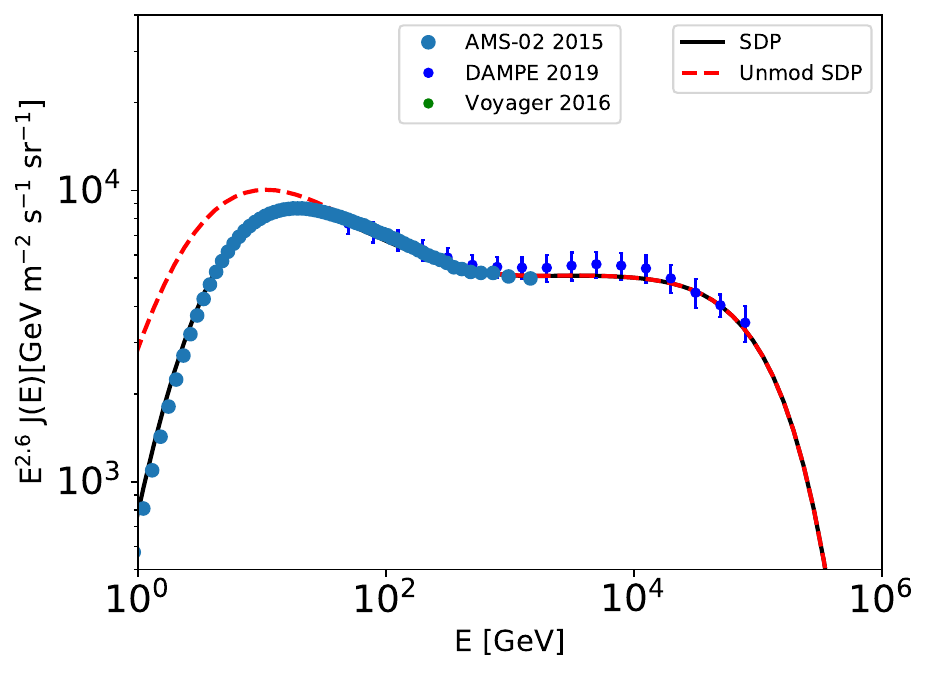}
    \caption{The CR proton spectrum is calculated using the CR SDP model and compared with observational data from AMS-02 \citep{2015PhRvL.114q1103A} and DAMPE \citep{2019SciA....5.3793A}. The red dashed line shows the spectrum without including solar modulation effects.}
    \label{fig5}
    \end{figure}

\begin{equation}
    \begin{aligned}
        \frac{\partial \psi(\vec{r}, p, t)}{\partial t}= & Q(\vec{r}, p, t)+\vec{\nabla} \cdot\left(D_{x x} \vec{\nabla} \psi-\vec{V}_c \psi\right) \\
        & +\frac{\partial}{\partial p}\left[p^2 D_{p p} \frac{\partial}{\partial p} \frac{\psi}{p^2}\right] \\
        & -\frac{\partial}{\partial p}\left[\dot{p} \psi-\frac{p}{3}\left(\vec{\nabla} \cdot \vec{V}_c\right) \psi\right]-\frac{\psi}{\tau_f}-\frac{\psi}{\tau_r}
        \end{aligned}
        \label{eq3}
\end{equation}
where $\psi(\vec{r}, p, t)$ represents the CR density per unit of total particle momentum $\rm p$ at position $\vec{r}$, $Q(\vec{r}, p, t)$ describes the source term, $\rm D_{x x}$ denotes the spatial diffusion coefficient, $\vec{V}_c$ is the convection velocity and $\rm \tau_f$ and $\rm \tau_r$ are the timescales for loss by fragmentation and radioactive decay, respectively. However, this work does not account for Galactic convection, so the convection velocity $\vec{V}_c$ is set to zero. The CR diffusion depends on the distribution of CR sources $\rm f(r,z)$, and the diffusion coefficient is described as\citep{2016ApJ...819...54G,2018PhRvD..97f3008G} 

\begin{table}[b]
\footnotesize
\centering
\caption{Parameters of the SDP model.}
\label{tab1}
\tabcolsep 3.5pt
\begin{tabular*}{0.47\textwidth}{lcccccc}    
\hline \hline
$D_0{ }\left[\mathrm{cm}^{2} \mathrm{~s}^{-1}\right]$ & $\delta_0$ & $N_m$ & $\xi$ & $\mathrm{n}$ & $v_A\left[\mathrm{~km} \mathrm{~s}^{-1}\right]$ & $z_0[\mathrm{kpc}]$ \\
\hline $4.5 \times 10^{28}$ & 0.64 & 0.24 & 0.1 & 4.0 & 6 & 4.5 \\
\hline 
\end{tabular*}
\end{table}

\begin{equation}
D_{x x}(r, z, \mathcal{R})=D_0 F(r, z) \beta^\eta\left(\frac{\mathcal{R}}{\mathcal{R}_0}\right)^{\delta_0 F(r, z)},
\label{eq4}
\end{equation}
where $\rm \delta_0 F(r, z)$ describes the turbulent characteristics of the local medium environment and $\rm D_0 F(r, z)$ represents the normalization factor of the diffusion coefficient at the reference rigidity.
\begin{equation}
    F(r, z)= \begin{cases}g(r, z)+[1-g(r, z)]\left(\frac{z}{\xi z_0}\right)^n, & |z| \leq \xi z_0 \\ 1, & |z|>\xi z_0\end{cases},
\end{equation}
here, $\rm \xi z_0$ denotes the half-thickness of the Galactic halo, and $\rm g(r, z)=N_m /[1+f(r, z)]$, where $\rm N_m$ is the normalization factor. The parameter $\rm n$ is used to describe the smoothness between the inner and outer halos, and the source distribution $\rm f(r,z)$ is a cylindrically symmetric continuous distribution\citep{1996A&AS..120C.437C,1998ApJ...509..212S,1998ApJ...504..761C}, 
\begin{equation}
f(r, z)=\left(\frac{r}{r_{\odot}}\right)^{1.25} \exp \left[-\frac{3.87\left(r-r_{\odot}\right)}{r_{\odot}}\right] \exp \left(-\frac{|z|}{z_s}\right),
\end{equation}
Where $\rm r_{\odot}$ = 8.5 kpc and $\rm z_s$ = 0.2 kpc. The Dpp describes the re-acceleration process of particles during propagation, and its coupling relationship with the spatial diffusion coefficient Dxx is given by:
\begin{equation}
D_{\mathrm{pp}} D_{\mathrm{xx}}=\frac{4 p^2 v_A^2}{3 \delta\left(4-\delta^2\right)(4-\delta)w}
\end{equation}
where $\rm v_A$ is the Alfvén speed, and $\rm w$ is the ratio of magneto-hydrodynamic wave energy density to the magnetic field energy density, which can be fixed to 1. Figures \ref{fig4} and \ref{fig5} present the comparison between the proton spectra and the B/C ratio obtained from observations and the predictions of the SDP model, respectively.

\end{appendices}

\bibliographystyle{aasjournal}
\bibliography{ref_v2}{}
\end{document}